\documentclass[twocolumn,showpacs]{revtex4}
\usepackage{graphicx}
\usepackage[]{epsfig}
\usepackage[T1]{fontenc}

\newcommand{\beq}{\begin{equation}}
\newcommand{\eeq}{\end{equation}}
\newcommand{\beqd}{\begin{displaymath}}
\newcommand{\eeqd}{\end{displaymath}}
\newcommand{\beqa}{\begin{eqnarray}}
\newcommand{\eeqa}{\end{eqnarray}}

\newcommand{\s}{\sigma}

\newcommand{\comment}[1]{}

\newcommand{\Tr}{{\rm Tr}\,}

\begin{document}
\title{Composite operators in Cubic Field Theories \\
and Link Overlap Fluctuations in Spin-Glass Models}
\author{Ada Altieri$^{1,2}$, Giorgio Parisi$^{1,3,4}$ and Tommaso Rizzo$^{5,1}$} 
\affiliation{$^1$ Dip. Fisica, Universit\`a "Sapienza", Piazzale A. Moro 2, I-00185, Rome, Italy  
\\
$^2$ LPTMS, CNRS, Univ. Paris-Sud, Universit\'e Paris-Saclay, 91405 Orsay, France
\\ $^3$ NANOTECH-CNR, UOS Rome, Universit\`a "Sapienza", PIazzale A. Moro 2,
I-00185, Rome, Italy 
\\$^4$ INFN, Piazzale A. Moro 2, 00185, Rome\\
$^5$ ISC-CNR, UOS Rome, Universit\`a "Sapienza", PIazzale A. Moro 2,
I-00185, Rome, Italy}

\begin{abstract}
We present a complete characterization of the fluctuations and correlations of the squared overlap in the Edwards-Anderson Spin-Glass  model in zero field.
The analysis reveals that the energy-energy correlation (and thus the specific heat) has a {\it different} critical behavior than the fluctuations of the link overlap in spite of the fact that the average energy and average link overlap have the same critical properties. More precisely the link-overlap fluctuations are larger than the specific heat according to a computation at first order in the $6-\epsilon$ expansion.
An unexpected outcome is that the link-overlap fluctuations have a subdominant power-law contribution characterized by an anomalous logarithmic prefactor which is absent in the specific heat.
In order to compute the $\epsilon$ expansion we consider the problem of the renormalization of quadratic composite operators in a {\it generic} multicomponent cubic field theory: the results obtained have a range of applicability beyond spin-glass theory.
\end{abstract}


\maketitle

\section{Introduction}

The critical properties of the Edwards-Anderson (EA) Spin-Glass model \cite{Edwards-Anderson} are the object of intense numerical study \cite{JanusII_2013,JanusII_preprint,Ballesteros00,Janus2010}.
Most results, including notably the existence of a phase transition in zero field at a finite temperature, have been established numerically.  In the course of the last decades many tools and observables have been introduced and in this paper we present an analytical investigation of a class of observables that are often measured in numerical simulations.

The order parameter of the model is the overlap between different replicas and other observables can be defined by taking powers of it.
Below the upper critical dimension the critical properties of these observables cannot be obtained from those of the order parameter and require the introduction of new critical exponents.  

In this paper we consider observables given by the products of two overlaps: in the case of Gaussian couplings, it is possible to prove by integration by parts that the local energy and its correlations correspond to appropriate combinations of overlaps.
When the order parameter is a single scalar the only observable of this kind is its square, but when the order parameter is a multi-index object we can define many different observables each characterized by a different critical exponent.
An important and much studied observable of this kind in the spin-glass context is the so-called link-overlap \cite{Katzgraber02}.
It is well-known that the average of the energy and the average of the link overlap have the same critical behavior and therefore one is naturally led to ask whether their respective fluctuations are also characterized by the same critical exponents or not. 
We will answer this question by considering the symmetries of the replicated Hamiltonian and showing that the products of the overlap can be classified in three orthogonal subspaces. The link-overlap and the energy belong to the same subspace (we will call it the {\it squared} subspace) but to different components. This implies our first result: the critical fluctuations of the energy and of the link-overlap are described by different critical exponents.
In order to obtain more information on the critical exponents we will determine them at first order in the $\epsilon=D_u-D$ expansion. From this it is difficult to obtain information for the exponents in $D=3$ which is rather faraway from the upper critical dimension $D_u=6$, nevertheless we obtain the qualitative information that {\it the link-overlap has larger fluctuations than the energy} near the upper critical dimension and this seems to remain true in $D=3$ according to preliminary numerical observations \cite{JanusII_preprint}.
A peculiar aspect is that the link-overlap fluctuations have also a subleading contribution characterized by an anomalous logarithmic prefactor which is missing in the specific heat.

We compute the $\epsilon$-expansion using standard field-theoretical methods \cite{Parisi88,LeBellac,Amit_MartinM,Collins84,ZinnJustin}.
Given a Landau theory for a second-order phase transition we consider the problem of renormalization of the corresponding field theory in the continuum limit and we extract the critical exponents from it. The products of the order parameters considered here are associated to the renormalization of the so-called composite operators (for this reason we will talk of composite operators also  for the observables in the original statistical mechanics context).
The field theory associated to the EA model in zero field is a multi-component cubic theory.
Thus we first consider the problem of renormalization of composite operators in the generic multi-component cubic theory obtaining the expressions for the renormalization constants at one-loop order.  Then we specialize these expressions to the special case of the EA model in zero field by imposing the corresponding additional symmetries.

The paper is organized as follows.
In section (\ref{squared}) we briefly introduce the EA model and the observables of interest.  We then study the properties of correlations of composite operators  and we diagonalize the corresponding correlation matrix using symmetry arguments.
In section (\ref{generic}) we discuss the computation of the $\epsilon$-expansion from the renormalization of the field theory. The subject is rather complex and covered in an extensive literature \cite{LeBellac,Amit_MartinM,Collins84,ZinnJustin}, thus we only illustrate in the simplest case of a scalar cubic field theory the procedure for extracting the critical exponents from the renormalized coupling constants. Finally we derive the formulas for a generic cubic theory.
In section (\ref{special}) we specialize the general expressions to the EA model obtaining the $\epsilon$-expansion for the fluctuations of the link-overlap. 
In section (\ref{conclusions}) we give our conclusions.

\section{Squared Overlaps in Spin-Glasses}
\label{squared}

We consider the following Edwards-Anderson Hamiltonian 
\beq
H=-\sum_{(x,y)}J_{xy}s_xs_y
\eeq
for a system of Ising spins defined on a $D$ dimensional Euclidean lattice with periodic boundary conditions. The quenched interactions are restricted on nearest-neighbors. The existence of a phase transition at a finite temperature and universality with respect to the distribution of the couplings have been established numerically \cite{Hasenbusch08}. 
The overlap field is defined as  
\begin{equation}
q^{ab}_{x}= \s^{a}_{x}\s^{b}_{x}
\end{equation}
where $a$ and $b$ label distinct replicas with the same realization of the random coupling.
Its spatial average gives the spin glass order parameter:
\begin{equation}
q^{ab}=\frac{1}{V} \sum_{x} q^{ab}_{x} \ ,
\end{equation}
where $V$ indicates the volume.

Let us consider now the spatial correlation function $G(r)$ and its Fourier transform $\chi(k)$: 
\begin{equation}
G(r)= \frac{1}{V} \sum_{x} \overline{ \left\langle q^{ab}_{x+r}q^{ab}_{x} \right\rangle} \hspace{0.1cm}
\ , \hspace{0.3cm}
\chi(k)=\frac{1}{V} \sum_{r} G(r) e^{ik \cdot r} \ .
\eeq
The fluctuations of the overlap $q^{ab}$ are directly connected to the spin glass susceptibility, which is an experimentally measurable quantity characterized by a sharp cusp at a finite temperature $T_c$ \cite{Gunnarsson91, Ballesteros00} and directly related to the non-linear magnetic susceptibility:
\beq
\chi_{SG}= \chi(0) \approx \chi_{nl} =V \overline {\left\langle q^2 \right\rangle} \ .
\eeq
In addition we want to consider the square of the overlap field in the same point, but since it takes a trivial value in the case of Ising spins we consider product of the overlaps at distance one on the lattice and this define the \emph{link-overlap field} $Q^{ab}_{x,\mu}$, which depends on the lattice site and the link direction $\mu$:
\beq
\label{link-overlap}
Q^{ab}_{x,\mu}=q^{ab}_{x}q^{ab}_{x+e_\mu} \ .
\eeq
Considering the Fourier transform of the deviation of the link overlap from its expectation value
\beq
\hat{Q}^{ab}_{k,\mu}=\frac{1}{V} \sum_{x}\delta Q^{ab}_{x,\mu} e^{i k \cdot x} \ ,
\eeq
we can define the \emph{link overlap susceptibility} evaluated at zero wave vector $\vec{k}$:
\beq
\chi_{link}(k,\mu)=V \overline{ \left\langle \vert \hat{Q}^{ab}_{k,\mu} \vert ^2 \right\rangle}
\eeq
\beq
\chi_{link}=\chi_{link}(0,\mu) 
\eeq
which is an interesting observable often measured in numerical studies.

The link overlap is an instance of a wider class of observables involving squares of the overlap field:
\beq
\hat{Q}^{ab,cd}_{x,\Delta}=q^{ab}_x q^{cd}_{x+\Delta}
\eeq
where $\Delta$ is a generic vector on the lattice
and we are interested in determining the spatial correlations of these quadratic observables:
\beq
\overline{\langle \hat{Q}^{ab,cd}_{x,\Delta} \hat{Q}^{a'b',c'd'}_{x',\Delta'}\rangle}-\overline{\langle  \hat{Q}^{ab,cd}_{x,\Delta} \rangle}\,\overline{\langle \hat{Q}^{a'b',c'd'}_{x',\Delta'} \rangle} \ .
\eeq
At the critical temperature, if we take the limit of 
$|x-x'|\rightarrow \infty$  at fixed finite $\Delta$ and $\Delta'$, these correlations have a power-law behavior as a function of $|x-x'|$ independently of $\Delta$ and $\Delta'$. However $\Delta=0$ is trivial for Ising spins and this is because one takes $\Delta$ as a unit vector on the lattice, {\it i.e.} the smallest non-trivial value.

In order to discuss the properties of the quadratic correlations it is useful to consider the corresponding Landau theory that will be essential in the following to evaluate the critical exponents.
Using standard arguments  one can argue that the critical properties of the Edwards-Anderson model  are determined by the following effective action defined on a replicated order parameter $q_{ab}(x)$ at zero field \cite{Harris76}:
\beq
H=\int d^Dx \left( {1 \over 4} \sum_{a\neq b} (\partial_\mu q_{ab})^2+{w_1 \over 6}\Tr q^3  \right) 
\label{HLaction}
\eeq
where $q_{ab}(x)$ is a real symmetric matrix and $a,b$ two replica indexes that can take values $1,2...n$. 
The order parameter is such that $q_{aa}=0$ and in the end we must take the quenched limit in which the number of replicas $n$ goes to zero.

If the distribution of the couplings is Gaussian, it is possible to obtain directly the Hamiltonian. In the generic case one can derive it under general symmetry considerations as usual in the Landau theory. Given the order parameter $q_{ab}(x)$ the action must indeed satisfy two constraints: the first one is obviously the replica symmetry, the second one is connected to the fact that the EA Hamiltonian in zero field is symmetric under reversal of the spins in each replica, therefore the replicated Hamiltonian must be invariant under the following transformation separately for each replica:
\beq
\{ q_{1a},q_{2a}, \dots, q_{na}  \} \rightarrow \{- q_{1a},-q_{2a}, \dots, -q_{na}  \} \ .
\label{Z2replicas}
\eeq
In the general case the RS theory has three quadratic terms and eight cubic terms but the above symmetry leads necessarily to the simpler Hamiltonian (\ref{HLaction}).

The quadratic observables defined above are naturally mapped into the connected correlations of the following composite operators:
\beq
\hat{Q}_{ab,cd}(x)\equiv q_{ab}(x)q_{cd}(x)
\eeq
\beq
G_{(ab,cd),(a'b',c'd')}(x,y)\equiv \langle  \hat{Q}_{ab,cd}(x)\hat{Q}_{a'b',c'd'}(y) \rangle_c
\label{G_correlations}
\eeq
Thus the critical behavior of the link overlap and of its fluctuations are identified with
\beq
\hat{Q}_{ab,ab}(x)\,,\ \langle  \hat{Q}_{ab,ab}(x)\hat{Q}_{ab,ab}(y) \rangle_c
\eeq 

In the case of Gaussian distribution of the couplings one can show explicitly that the averages of the energy and of the link-overlap have the same critical behavior, this result however is more general and can be derived from the replicated action above.  
Indeed standard arguments lead to identify the critical behavior of the energy with the one of the quadratic term in the Hamiltonian
\beq
E = {1 \over n}\sum_{ab} \langle q_{ab}^2(x) \rangle \ .
\eeq
Therefore the critical behavior of the average value of the energy can be identified with that of the link overlap, while at the level of fluctuations we will see that they are different.
In order to clarify the difference we introduce the following quantities that will be discussed extensively in the following: 
\beq
G_1^{(2)}(x-y) \equiv \langle q_{ab}^2(x)q_{ab}^2(y)\rangle-
 \langle q_{ab}^2(x) \rangle \langle  q_{ab}^2(y)\rangle
\eeq
\beq
G_2^{(2)}(x-y) \equiv \langle q_{ab}^2(x)q_{ac}^2(y)\rangle-
 \langle q_{ab}^2(x) \rangle \langle  q_{ac}^2(y)\rangle
\eeq
\beq
G_3^{(2)}(x-y) \equiv \langle q_{ab}^2(x)q_{cd}^2(y)\rangle-
 \langle q_{ab}^2(x) \rangle \langle  q_{cd}^2(y)\rangle
\eeq
We see that the fluctuations of the link-overlap must be naturally identified with  $G_1^{(2)}(x-y)$. On the contrary, the fluctuations of the energy  receive three different contributions and they depend also on off-diagonal terms. The energy-energy correlation function is indeed:
\begin{displaymath}
\left\langle E(0)E(x) \right\rangle_{c} = {1 \over n}\left\langle \sum_{ab}q_{ab}^2(0)\sum_{cd} q^2_{cd}(x) \right\rangle_{c} =
\end{displaymath}
\beq
=G^{(2)}_{1}(x)+ 2(n-2) G^{(2)}_{2}(x) + \frac{(n-2)(n-3)}{2} G^{(2)}_{3}(x) 
\label{energy}
\eeq
As we will see at the end of the next subsection the above expression  for the energy-energy correlations corresponds to the so-called squared-longitudinal eigenvalue $g_{L^{(2)}}(x)$ and therefore it has a different scaling behavior from the fluctuations $G^{(2)}_{1}(x)$ of the link overlap that receives a contribution also from another (dominant) eigenvalue.

\subsection{The Structure of Correlations of Quadratic Observables}

In order to characterize the behavior of the correlation function (\ref{G_correlations}) we must find a basis for the space of the observables $\hat{Q}_{ab,cd}(x)$ where it is diagonal. Indeed at the critical point each diagonal component will be characterized by a power-law decay with a given exponent.
Now in order to proceed it is useful to consider the implications of the invariance of the replicated Hamiltonian with respect to the symmetry (\ref{Z2replicas}). The main consequence is that a generic correlation function will be zero unless each distinct replica index appears an even number of times.
Immediate implications for two-point functions of the order parameter are:
\beq
G_1(x,y)\equiv \langle q_{ab}(x)q_{ab}(y) \rangle \neq 0\, ,
\eeq
\beq
 G_2(x,y)\equiv \langle q_{ab}(x)q_{ac}(y) \rangle = 0\, ,
\eeq
\beq
G_3(x,y)\equiv \langle q_{ab}(x)q_{cd}(y) \rangle = 0\, .
\eeq
Similarly one can show that the only cubic correlation that does not vanish is
\beq
\langle q_{ab}(x)q_{bc}(y)q_{ac}(z)\rangle\neq 0 \ .
\eeq
The invariance can be also used to understand the structure of quadratic composite operators.
In particular we want to determine which elements of $G_{(ab,cd),(a'b',c'd')}(x,y)$ are non-zero.
We can distinguish three types of subspaces:

\begin{itemize}

\item
 if $(ab,cd)$ are all different then $(a'b',c'd')$ must be equal to  $(ab,cd)$ or a permutation. In other words, given four different indexes, say $1,2,3,4$, the three composite operators
\beq
q_{12}(x)q_{34}(x)\, , q_{13}(x)q_{24}(x)\, , q_{14}(x)q_{23}(x)\, ,
\eeq  
form a closed sub-space. Replica symmetry implies that the elements of the corresponding $3\times 3$ interaction  matrix can take only two values (diagonal or off-diagonal) and therefore we have one eigenvalue (the symmetric one) with degeneracy one and another eigenvalue (the antisymmetric one) with degeneracy two. The number of such three-dimensional subspaces is given by the possible choice of four different indexes: $n(n-1)(n-2)(n-3)/4!$ .

\item
If $(ab,cd)$ are such that there are two and only two equal indexes, (say we consider the operator $q_{12}(x)q_{13}(x)$), the unmatched indexes ($2$  and $3$) must be matched by  $(a'b',c'd')$, {\it i.e.} we must have $q_{j2}(y)q_{j3}(y)$ where $j$ can be either equal to $1$ or not.
This implies that the couple of unmatched indexes $2,3$ in $q_{j2}(x)q_{j3}(x)$ define a closed $n-2$ dimensional subspace where each element corresponds to a value of $j$. For symmetry reasons the interaction matrix can take only two values, (diagonal or off-diagonal), therefore we have the symmetric eigenvalue with multiplicity one and the antisymmetric eigenvalue with multiplicity $n-3$.
The total number of subspaces is given by the couples of different indexes: $n(n-1)/2$.

\item
Lastly if we have two couples of equal indexes in the left, {\it e.g.} in $q_{12}^2(x)$, we must also have two couples of equal indexes on the right, {\it i.e.} $q_{ab}^2(y)$ where $ab$ can be any couple of indexes. As a consequence the structure of this subspace (that we call quadratic in the following) is similar to that of $q_{ab}$ and we have accordingly three eigenvalues: longitudinal, anomalous and replicon. A detailed analisys of this subspace, that we call the squared subspace, will be given in the following subsection.

\end{itemize}

\subsubsection{General Properties of the Squared Subspace}

We now present a more detailed analysis of the squared subspace, {\it i.e.} the set of composite operators of the form $q_{ab}^2(x)$.
This is a rather important subspace because as we have seeen before both the energy and the link-overlap  belong to it.
Replica-Symmetry group acts on this subspace in a way formally equivalent to that of the order parameter $q_{ab}(x)$ and therefore its diagonalization is discussed extensively in the literature \cite{Almeida,Temesvari02,DeDominicis06,DeDominicis_field}.

As we already said RS implies that the correlations of the squared order parameters can take only the three possible values $G_1^{(2)}(x-y)$, $G_2^{(2)}(x-y)$ and $G_3^{(2)}(x-y)$, defined above, depending on how many indexes are different.  
Using the classic Almeida-Thouless results (see {\it e.g.} \cite{Temesvari02}) we can express the three susceptibilities in terms of three diagonal propagators that we call the squared-longitudinal, squared-anomalous and squared-replicon subspaces in agreement with the standard notation.
We have:
\beqa
G_1^{(2)}(x) &  = & {2 \,g_{L^{(2)}}(x) \over n(n-1)}+{2\, g_{A^{(2)}}(x) \over n}+{(n-3)\, g_{R^{(2)}}(x) \over n-1}
\nonumber
\\
G_2^{(2)}(x) & = &  {2\, g_{L^{(2)}}(x) \over n(n-1)}+{(n-4)\, g_{A^{(2)}}(x) \over n(n-2)}+{(3-n)\, g_{R^{(2)}}(x) \over n^2-3 n +2}
\nonumber
\\
G_3^{(2)}(x) &  = & {2 \,g_{L^{(2)}}(x) \over n(n-1)}-{4 \,g_{A^{(2)}}(x) \over n(n-2)}+{2\, g_{R^{(2)}}(x) \over n^2-3 n +2}
\eeqa 
At finite $n$ each of the three diagonal components has a power-law decay at criticality characterized by a different exponent:
\beqa
g_{R^{(2)}}(x) &  \propto & {1 \over x^{2D-4-2\eta_{R^{(2)}}}}
\nonumber
\\
g_{L^{(2)}}(x) &  \propto & {1 \over x^{2D-4-2\eta_{L^{(2)}}}}
\nonumber
\\
g_{A^{(2)}}(x) &  \propto & {1 \over x^{2D-4-2\eta_{A^{(2)}}}} \ .
\eeqa
However in the $n \rightarrow 0$ limit there is a peculiar change due to the fact that the longitudinal and anomalous correlations (and the corresponding exponents $\eta_{L^{(2)}}$ and $\eta_{A^{(2)}}$) are equal. This on the other hand guarantees that the above expressions for $G_1^{(2)}(x)$, $G_2^{(2)}(x)$, and $G_3^{(2)}(x)$ are not singular at $n=0$ as it may appear.
In order to carefully study the problem we need to consider the difference between the longitudinal and anomalous propagator divided by $n$ in the $n \rightarrow 0$ limit:
\beq
\Delta g(x) \equiv \lim_{n \rightarrow 0} { g_{A^{(2)}}(x)-g_{L^{(2)}}(x) \over n} \ .
\eeq 
We can then take the quenched limit  and obtain:
\beqa
G_1^{(2)}(x) &  = & 2 \Delta g (x)- 2 g_{L^{(2)}}(x) +3 g_{R^{(2)}}(x) 
\nonumber
\\
G_2^{(2)}(x) & = & 2 \Delta g (x)- { 3 \over 2} g_{L^{(2)}}(x)+{3 \over 2} g_{R^{(2)}}(x) 
\nonumber
\\
G_3^{(2)}(x) &  = & 2 \Delta g (x)-  g_{L^{(2)}}(x)+ g_{R^{(2)}}(x) 
\eeqa
Quite interestingly the asymptotic behavior of $\Delta g(x)$ has an additional logarithmic prefactor to the power-law that results from the derivative of the finite-$n$ power-law behaviors: 
\beqa
g_{R^{(2)}}(x) &  \propto & {1 \over x^{2D-4-2\eta_{R^{(2)}}}}
\\
g_{L^{(2)}}(x) &  \propto & {1 \over x^{2D-4-2\eta_{L^{(2)}}}}
\\
\Delta g(x) &  \propto & {\Delta \eta_2\,\ln x \over x^{2D-4-2\eta_{L^{(2)}}}} \ .
\eeqa
Note that the above mechanism leading to logarithmic prefactors has been discussed earlier by Cardy as a rather general feature of replicated theories \cite{Cardy99}. 

The logarithmic corrections have a numerical prefactor proportional to the difference between the degenerate eigenvalues divided by $n$: 
\beq
\Delta \eta_2 \equiv \lim_{n \rightarrow 0} { \eta_{A^{(2)}}-\eta_{L^{(2)}} \over n} \ .
\eeq
The presence of the logarithmic corrections would be masked in $G_1^{(2)}(x)$,  $G_2^{(2)}(x)$, $G_3^{(2)}(x)$ if the squared-replicon is dominant, {\it i.e.} if $\eta_{R^{(2)}}>\eta_{L^{(2)}}$. As we will see in the next section this is actually the case at least at first order in the $\epsilon$ expansion.
In order to observe numerically the different exponents and the logarithmic scaling one should study appropriate combinations of $G_1^{(2)}(x)$,  $G_2^{(2)}(x)$ and $G_3^{(2)}(x)$ that can be obtained inverting the above $n=0$ expressions:
\beqa
 g_{R^{(2)}}(x) &  = & G_1^{(2)}(x)- 2 G_2^{(2)}(x) + G_3^{(2)}(x)
\\
 g_{L^{(2)}}(x) &  = & G_1^{(2)}(x)- 4 G_2^{(2)}(x) + 3 G_3^{(2)}(x)
\\
\Delta g (x) &  = & { 3 \over 2}G_3^{(2)}(x)-  G_2^{(2)}(x) \ .
\eeqa
It is important to observe that the above expression for the squared-longitudinal eigenvalue coincides with  that of the energy-energy correlations (see eq. (\ref{energy}) above). This implies that, {\it at variance with the link-overlap fluctuations  $G_1^{(2)}(x)$, the energy-energy correlations do not depend on the dominant replicon eigenvalue $g_{R^{(2)}}(x)$ and they are thus subdominant.}

The above identification of the energy-energy correlations with $ g_{L^{(2)}}(x)$ implies also that 
 the logarithmic corrections are absent in the energy-energy correlations while they would be subdominant in the link-overlap fluctuations, therefore in order to observe them one should consider the above combination for $\Delta g(x)$. 
 
The above structure was derived by symmetry arguments: it is interesting to check that it holds  at all orders in the loop expansion of the replicated action (\ref{HLaction}). 
In order to compute the loop expansion one has to attach a couple of different replica indexes to the Feynman diagrams of the scalar cubic theory. The vertex $w_1$ is such that two indexes entering from the same leg must get out from different legs.
This has two consequences: i) each index coming from one of the external legs must exit at a different leg through a simple path (no bifurcations or crossings); ii) internal indexes to be summed upon must instead run on closed paths. This implies that at all orders: i) the propagator remains diagonal upon renormalization and ii) no additional cubic vertexes besides $w_1$ are generated by the loop expansion.
Similarly one can show that the decomposition of $G_{(ab,cd),(a'b',c'd')}(x,y)$ derived above is satisfied at all orders as it should.

\section{The generic multi-component cubic theory}
\label{generic}

In the following we consider the problem of the renormalization of the generic cubic field theory and then we obtain expressions for the critical exponents of the corresponding second-order phase transition.
The connection between the problem of the continuum limit in field theories (which is solved by renormalization) and the problem of critical phenomena is discussed at length in many places \cite{Parisi88,LeBellac,Amit_MartinM,Collins84,ZinnJustin}  and we will not discuss it here.
Instead we recall the procedure in the simplest case of a scalar massless cubic theory and then we repeat the same steps in a completely general context. In the next section we specialize again, this time considering the replicated theory of the spin-glass in zero field.

\subsection{The scalar massless theory}

The massless scalar cubic field theory is given by \cite{LeBellac,Amit_MartinM,Collins84}:
\beq
H=\int d^Dx \left[ {1 \over 2} (\partial_\mu \phi_0)^2+{1 \over 6} u \, \phi_0^3  \right]  \ .
\eeq
The above expression written in terms of renormalized variables reads:
\beq
H=\int d^Dx \left[ {1 \over 2} Z(\partial_\mu \phi)^2+{1 \over 6} g\, k_D^{-1/2} \mu^{\epsilon/2} Z_g \, \phi^3  \right] 
\eeq
where $\epsilon=6-D$, $\mu$ is the momentum scale and
\beq
\phi=Z^{-1/2}\phi_0 
\eeq
is the renormalized field.
Besides, $k_D$ is a factor introduced for later convenience:
\beq
k_D={S_D \over (2 \pi)^D}
\eeq
where $S_D$ is the surface of the unit sphere in $D$ dimensions.

The renormalization constants are determined in the minimal subtraction scheme imposing that the renormalized proper vertexes are finite:
\beqa
\Gamma^{(2)}_r & = & Z \Gamma^{(2)}
\\
\Gamma^{(3)}_r & = & Z^{3/2} \Gamma^{(3)}
\\
\Gamma^{(1,2)}_r & = & \zeta \, Z \Gamma^{(1,2)}
\eeqa
where 
\beq
\Gamma^{(1,2)}={d \over d \, m_0} \Gamma^{(2)} 
\eeq
and $m_0$ is  a mass source in the bare Hamiltonian:
\beq
\delta H=\int d^Dx  {1 \over 2} m_0 \phi_0^2 \ .
\eeq
We consider as usual the following conditions that are consistent with the tree level:
\beq
Z=1+O(g^2),\  Z_g=1+(g^2),\ \zeta=1+O(g^2)\,.
\eeq
The expression for the $\beta$-function $\beta(g)$ can be obtained by derivation of the following equation with respect to $\mu$ at $u$ fixed:
\beq
k_D^{1/2} \, u = \mu^{\epsilon/2}G, \ \ \ G(g)\equiv g Z_g/Z^{3/2}
\eeq
and leads to
\beq
\beta(g)=-{\epsilon \over 2}G \left({d G \over dg}\right)^{-1} \ .
\eeq
Introducing the functions
\beq
\eta(g)\equiv \beta(g){d \ln Z \over dg} \ ,
\eeq
\beq
\eta_2(g)\equiv \beta(g){d \ln \zeta \over dg} \ ,
\eeq
the critical exponents are given by $\eta=\eta(g^*)$ and $\eta_2=\eta_2(g^*)$ where $g^*$ is the critical value of the renormalized coupling constant identified by $\beta(g^*)=0$.
The exponent $\eta_2$ determines the anomalous dimension of the product through:
\beq
d_{\phi^2}=D-2-\eta_2 \ ,
\eeq
therefore the correlation length exponent is:
\beq
d_{\phi^2}=D-{1 \over \nu} \ \rightarrow  \hspace{0.3cm} \nu={1 \over 2+\eta_2} \ .
\eeq
\subsubsection{The computation of the coupling constants}

At one loop order we obtain for the renormalized constants:
\beq
Z=1-{g^2 \over 6 \epsilon}\ .
\eeq
\beq
Z_g=1-{g^2 \over \epsilon} \hspace{0.2cm} \ , \hspace{0.2cm} G=g-\frac{3g^3}{4\epsilon}
\eeq
The $\beta$ function in our case reads:
\beq
\beta(g)=-{1 \over 2}\epsilon\, g-{3 \over 4}g^3\ .
\eeq
Including a $\phi^2$ insertion and using the condition that  $Z \zeta \Gamma^{(1,2)}$ is finite (where $\Gamma^{(1,2)}$ is the generating functional with two external legs evaluated in presence of an insertion) we obtain:
\beq
\zeta=1-{5 \over 6 \epsilon}g^2
\eeq
which implies at the leading order:
\beq
\eta(g)={g^2 \over 6} \hspace{0.2cm} \ , \hspace{0.3cm} \eta_2(g)={ 5 \over 6}g^2 \ .
\eeq

\subsection{The generic massless theory}

We now proceed to the generalization of the above formulas.
The general action is
\beq
H=\int d^Dx \left[ {1 \over 2} \sum_i(\partial_\mu \phi_{0,i})^2+{1 \over 6} \sum_{ijk} u_{ijk} \, \phi_{0,i} \phi_{0,j} \phi_{0,k} \right] 
\eeq
where $u_{ijk}$ is symmetric in its three indexes and the generic index $i$ can take values from $1$ up to $M$.

We want to introduce a renormalized coupling constant as:
\beq
k_D^{1/2}u_{ijk}=\mu^{\epsilon/2}g_{ijk}+O(g^3)
\eeq
and we renormalize the fields according to 
\beq
\phi_{i}=(Z^{-1/2})_{ij}\phi_{0,j}
\eeq
(above and in the following we use the notation for which repeated indexes are summed upon) with 
\beq
Z_{ij}=\delta_{ij}+O(g^2) \ .
\eeq

The renormalized vertexes are defined as \cite{ZinnJustin}:
\beqa
\Gamma^{(2)}_{r,ij} & = & (Z^{1/2})_{ii'} (Z^{1/2})_{jj'}\Gamma^{(2)}_{i'j'}
\nonumber
\\
\Gamma^{(3)}_{r,ijk} & = & (Z^{1/2})_{ii'} (Z^{1/2})_{jj'}(Z^{1/2})_{kk'}\Gamma^{(3)}_{i'j'k'}
\\
\Gamma^{(1,2)}_{r,(ij),(kl)} & = & \zeta_{(ij),(i'j')} \, (Z^{1/2})_{kk'}(Z^{1/2})_{ll'} \Gamma^{(1,2)}_{(i'j'),(k'l')}  \ .
\nonumber
\eeqa
The function $u(g)$, from which the generalized $\beta$ function can be obtained, and the functions $Z_{ij}$ and $\zeta_{(ij),(i'j')}$ are determined in the minimal scheme imposing the finiteness of the renormalized vertexes with:
\beq
\zeta_{(ij),(i'j')}={1 \over 2}(\delta_{ii'}\delta_{jj'}+\delta_{ij'}\delta_{ji'})+O(g^2) \ .
\eeq
The diagrams are the same of the scalar theory.

For the field renormalization we have:
\beq
\Gamma^{(2)}_{ij}=\mu^2\left(\delta_{ij}+{1 \over 6 \epsilon}g_{imn}g_{mnj}\right)
\eeq
from which we obtain:
\beq
Z_{ij}=\delta_{ij}-{1 \over 6 \epsilon}A_{ij}
\eeq
with
\beq
A_{ij}\equiv g_{imn}g_{mnj} \ .
\eeq
For the coupling constant renormalization we have:
\beq
\Gamma^{(3)}_{ijk}=u_{ijk}+u_{ilm}u_{jln}u_{knm} k_D {\mu^{-\epsilon} \over \epsilon}
\eeq
from which:
\begin{displaymath}
k_D^{1/2}\Gamma^{(3)}_{r,ijk}  =  k_D^{1/2} (Z^{1/2})_{ii'} (Z^{1/2})_{jj'}(Z^{1/2})_{kk'}\Gamma^{(3)}_{i'j'k'}=
\end{displaymath}
\begin{displaymath}
 =  -{\mu^{\epsilon/2}\over 12 \epsilon}A_{ii'}g_{i'jk}
-{\mu^{\epsilon/2}\over 12 \epsilon}A_{jj'}g_{ij'k}
-{\mu^{\epsilon/2}\over 12 \epsilon}A_{kk'}g_{ijk'}+
\end{displaymath}
\begin{displaymath}
+k_D^{1/2} u_{ijk}+B_{ijk} {\mu^{\epsilon/2} \over \epsilon}
\end{displaymath}
with
\beq
B_{ijk} \equiv g_{ilm}g_{jln}g_{knm} \ .
\eeq
The condition that the above expression remains finite implies that:
\begin{displaymath}
k_D^{1/2} u_{ijk}= \mu^{\epsilon/2} \left[ g_{ijk}-B_{ijk} {1 \over \epsilon}+ \right.
\end{displaymath}
\begin{displaymath}
\left.+{1\over 12 \epsilon}A_{ii'}g_{i'jk}
+{1\over 12 \epsilon}A_{jj'}g_{ij'k}
+{1\over 12 \epsilon}A_{kk'}g_{ijk'} \right]
\end{displaymath}
from which  differentiating with respect to $\mu$ at $u_{ijk}$ fixed and solving recursively we obtain the expression for the generalized $\beta$-function:
\begin{displaymath}
\beta_{ijk}=-{\epsilon \over 2}g_{ijk}-B_{ijk}+
\end{displaymath}
\begin{displaymath}
+{1\over 12 }A_{ii'}g_{i'jk}
+{1\over 12 }A_{jj'}g_{ij'k}
+{1\over 12 }A_{kk'}g_{ijk'} \ .
\end{displaymath}
The dimension of the fields can be extrapolated from the eigenvalues of the following operator that appears in the Callan-Symanzik equation:
\beq
\eta_{ij}= 2\beta_{i'j'k'} \left({d Z^{1/2} \over d g_{i'j'k'}} Z^{-1/2} \right)_{ij}\ .
\eeq
Finally, we obtain:
\beq
\eta_{ij}={1 \over 6}g_{imn}g_{mnj} \ .
\eeq
For the operator insertion the scalar expression generalizes to:
\begin{equation}
\Gamma^{(1,2)}_{(ij),(kl)}={1 \over 2}(\delta_{ik}\delta_{jl}+\delta_{il}\delta_{jk})
+{1 \over 2 \epsilon} (g_{ikn}g_{njl}+g_{iln}g_{jkn})
\end{equation}
from which the operator renormalization can be obtained as:
\begin{displaymath}
\zeta_{(ij),(kl)}={1 \over 2}(\delta_{ik}\delta_{jl}+\delta_{il}\delta_{jk})+
\end{displaymath}
\begin{displaymath}
+{1\over 24 \epsilon}\delta_{ik}A_{jl}
+{1\over 24 \epsilon}\delta_{il}A_{jk}
+{1\over 24 \epsilon}\delta_{jk}A_{il}
+{1\over 24 \epsilon}\delta_{jl}A_{ik}+
\end{displaymath}
\begin{displaymath}
-{1 \over 2 \epsilon} (g_{ikn}g_{njl}+g_{iln}g_{jkn}) \ .
\end{displaymath}
The dimension of the composite operator derives from the following operator that also appears in the  Callan-Symanzik equation:
\beq
\eta^{(2)}_{ij,kl}=\beta_{i'j'k'} \left({d \zeta \over d g_{i'j'k'}} \zeta^{-1} \right)_{ij,kl}
\eeq
where $\zeta_{ij,kl}$ defined above is considered as a linear operator restricted to the space of symmetric matrices $A_{ij}$ with the scalar product $A \cdot B \equiv\sum_{ij}A_{ij}B_{ij}$. 

Note that on this space the $O(1)$ term $(\delta_{ik}\delta_{jl}+\delta_{il}\delta_{jk})/2$ in $\zeta$ is equivalent to the identity. Correspondingly the dimension of the composite operator is given by the $M(M+1)/2$ ($M$ is the number of components of the field) eigenvalues of $\eta^{(2)}_{ij,kl}$ (corresponding to symmetric eigenvectors).
We finally arrive at:
\begin{displaymath}
\eta^{(2)}_{(ij),(kl)}=
-{1\over 24}\delta_{ik}A_{jl}
-{1\over 24}\delta_{il}A_{jk}
-{1\over 24}\delta_{jk}A_{il}
-{1\over 24}\delta_{jl}A_{ik}+
\end{displaymath}
\begin{displaymath}
+{1 \over 2} (g_{ikn}g_{njl}+g_{iln}g_{njk}) \ .
\end{displaymath}
Typically the energy is associated to the most symmetric composite operator $\sum_i \phi_{i,0}^2$. Calling $\eta_2$ the corresponding eigenvalue we find:
\beq
\nu=(2+\eta_2)^{-1}\, , \ \ \alpha=2 -\nu D
\eeq
where as usual $\nu$ and $\alpha$ are the critical exponents associated to the correlation length and to the specific heat.
The critical behavior with the temperature of the fluctuations of a composite operator associated to a different eigenvalue $\eta_2'$ of the matrix $\eta^{(2)}_{(ij),(kl)}$  are then given by
\beq
\alpha'=\alpha+2\, \nu\,(\eta_2'-\eta_2)  \ .
\eeq

\section{Critical Exponents of the square subspace: Specific heat and Link-overlap fluctuations}
\label{special}

In this section we apply the generic formulas derived above to the replicated theory (\ref{HLaction}) in order to compute the dimension  of the composite operators in the squared subspace,  we will thus determine $\eta_{L^{(2)}}$, $\eta_{A^{(2)}}$ and $\eta_{R^{(2)}}$ to the first order in the $\epsilon$ expansion.

In the previous section we have considered the general case where the cubic interaction term is given by a generic tensor $u_{ijk}$. Special cases are those where the tensor possesses some symmetry that reduces the number of its independent components. Note that the symmetry is preserved at all orders in the perturbation expansion. An important subclass is when there are enough symmetries in the Hamiltonian to determine the tensor up to a single coupling constant, in this case we can write $u_{ijk}=u d_{ijk}$ and $g_{ijk}=g d_{ijk}$ where $d_{ijk}$ is one instance of the symmetric tensor.
An important instance of this subclasss is the percolation problem when mapped to a Potts model.
The replicated theory (\ref{HLaction}) also belongs to this class because, as we have seen before, the RS symmetry and the $Z_2$ symmetry imply that there is only one type of cubic term with coupling constant $w_1$.
A detailed third-order treatment of the problem for a generic symmetric tensor $d_{ijk}$ has been given in \cite{DeAlcantara81} and the results have  been applied to the above spin-glass Hamiltonian in \cite{Green85} where the third order $\epsilon$-expansions for the critical exponents $\nu$ and $\eta$ are presented (errors in \cite{Green85} have been corrected in \cite{Moore-Yeo}). We note that it is not clear at present if these series can be resummed in dimension $D=3$ (see also \cite{Elderfield-McKane}).

We first rederive  the results of \cite{DeAlcantara81}  at the lowest order using the results of the previous section. In this case clearly the indexes $i,j,k,l$ run from one to  $n(n-1)/2$ (the number of independent components of the symmetric matrix $Q_{ab}$).
For the $\beta$ function  we have:
\beq
B_{ijk}=\hat{\beta} g^3 d_{ilm}d_{jln}d_{knm}=\hat{\beta} g^3 d_{ijk}\ , \hspace{0.2cm} \hat{\beta}=1+(n-3)
\label{Bijk}
\eeq
where the $n-3$ comes from the case in which each entering replica index exits at the first possible vertex, while the 1 comes from the case in which each replica index crosses a vertex and then exits at the next \cite{Green85}. 
Note that  $ d_{ilm}d_{jln}d_{knm}$ is proportional to $ d_{ijk}$ because we are working in the special case \cite{DeAlcantara81} where the symmetries of the problem determines each third-rank tensor up to a multiplicative factor.
The two-point function remains diagonal with:
\beq
A_{ij}=\hat{\alpha} g^2 \delta_{ij} \ , \hspace{0.3cm} \hat{\alpha}=2(n-2) 
\eeq
(we are following the conventions of Ref. (\cite{DeAlcantara81}) and thus $\hat{\alpha}$ and $\hat{\beta}$ in the above formulas should not be confused with the specific heat exponent and the $\beta$ function).
One then obtains:
\beq
\beta(g)=-{\epsilon \, g\over 2}-\hat{\beta} g^3+{\hat{\alpha} \over 4}g^3
\eeq
from which the fixed point reads:
\beq
g_*^2={2 \epsilon \over \hat{\alpha}-4 \hat{\beta}}={\epsilon \over 2 -n}
\eeq
and the anomalous dimension of the overlap is:
\beq
\eta={g^2 \hat{\alpha} \over 6}={\hat{\alpha} \epsilon/3 \over \hat{\alpha}-4 \hat{\beta}}=-{\epsilon \over 3} \ .
\eeq
As we have seen in the previous section the dimension of the composite operator is associated to the eigenvalues of the matrix:
\begin{equation}
\eta^{(2)}_{(ij),(kl)}=-{1\over 12}(\delta_{ik}\delta_{jl}+\delta_{il}\delta_{jk})\alpha g^2+{1 \over 2} (g_{ikn}g_{njl}+g_{iln}g_{njk})
\end{equation}
where we have used $A_{ij}=\hat{\alpha} g^2 \delta_{ij}$.
Projecting on the squared subspace means computing the eigenvalues of $\eta^{(2)}_{(ii),(kk)}$. The indexes $i$ and $k$ label the elements of the matrix $Q_{ab}$, as a consequence $\eta^{(2)}$ restricted to the squared subspace has the form of the generic mass matrix in RS theories \cite{Temesvari02} $M_{(ab),(cd)}$. Just for reader convenience we recall that such a matrix can assume three possible values:
\beq
M_1=M_{(ab),(ab)}\, , M_2=M_{(ab),(ac)}\, ,M_3=M_{(ab),(cd)} 
\eeq
and it has three eigenvalues called longitudinal, anomalous and replicon:
\beqa
r_R & = & M_1-2 M_2+M_3 \, ,
\\
r_A & = & M_1+(n-4)M_2-(n-3)M_3\, ,
\\
r_L & = & M_1 + 2(n-2)M_2+{(n-2)(n-3)\over 2} M_3
\eeqa
where in the $n \rightarrow 0$ limit longitudinal and anomalous are degenerate:
\beq
r_A=r_L=M_1-4M_2+3 M_3 \ .
\eeq
One can see that the $\eta^{(2)}_{(ii),(kk)}$ matrix is given by:
\begin{displaymath}
\eta^{(2)}_{(ii),(kk)}=-{1\over 6}\delta_{ik}\alpha g^2+g_{ikn}g_{nik}
\end{displaymath}
where the first term contributes to $M_1$, while the second one contributes to $M_2$:
\beq
M_1=-{n-2 \over 3} g^2\, , M_2=g^2\, ,M_3=0 
\eeq
(one can check that $M_3$ becomes non zero at the next order in the loop expansion).
Note that we are borrowing definitions from the case of the spin-glass in a field theory where $M_{(ab),(cd)}$ is associated to the correlations of the overlap and not as in our case to the overlap squared. 

It is clear that one should identify $\eta_{L^{(2)}}$ with $r_L$ and $\eta_{R^{(2)}}$ with $r_R$ as given by the previous formulas and  therefore the anomalous dimensions computed at the fixed point $g_*$ are: 
\beq
\eta_{L^{(2)}}=-{5 \over 3}\epsilon\, , \ \ \eta_{R^{(2)}}={n+4 \over 3(n-2)}\epsilon=-{2 \over 3}\epsilon + O(n) .
\eeq
We see that the square-replicon is larger than the squared-longitudinal and therefore the link overlap has larger fluctuations then the energy.
The correlation length exponent $\nu$ is then given by:
\beq
\nu=(2+\eta_{L^{(2)}})^{-1}={1 \over 2}+{5 \over 12} \epsilon
\eeq
in agreement with earlier results \cite{Harris76,Green85}. 

The specific heat behavior is given by
\beq
\alpha=-1-2 \epsilon
\eeq
while the corresponding exponent for the link overlap is larger:
\beq
\alpha_{link}=-1- \epsilon \ .
\eeq
Note that a divergence of the link-overlap fluctuations corresponds by definition to a positive $\alpha_{link}$.
Since the mean-field value ($\epsilon=0$) is negative it is not possible to observe such a change in perturbation theory. Nevertheless a positive value of the coefficient of the $O(\epsilon)$ correction would make a change of sign more likely. Instead the coefficient of the $O(\epsilon)$ term is negative and the only qualitative information we get is that it is smaller in absolute value than that of the specific heat, implying that the link-overlap fluctuations are larger than the specific heat.

\section{Conclusions}
\label{conclusions}

We have illustrated a characterization of the fluctuations and correlations of quadratic observables in the Edwards-Anderson spin-glass model in zero field.
Notably our study reveals that the energy-energy correlations (thus the specific heat) have a different critical behavior than the fluctuations of the link overlap. According to a computation at first order in the $\epsilon$ expansion the link-overlap fluctuations are larger than the specific heat.
This has to be contrasted with the well-known result that the average energy and link overlap have the same critical behavior.

In order to compute the $6-\epsilon$ expansion we approached the problem in terms of the renormalization of quadratic composite operators in a {\it generic} multicomponent cubic field theory obtaining results that have a wider range of applicability than spin-glass theory.
An unexpected outcome is that some critical contributions to the link-overlap fluctuations have a logarithmic prefactor absent in the specific heat.
An interesting problem is that this anomalous logarithmic prefactor is associated to a subdominant term (that we called square-longitudinal) and thus one should devise an appropriate procedure  to observe it in a numerical simulation.

\section*{Acknowledgments}
We would like to thank the Janus group for sharing information on the behavior of the link overlap and its correlations.

\end{document}